\documentclass[%
 aip,
 amsmath,amssymb,
 reprint,%
]{revtex4-1}

\usepackage{graphicx}
\usepackage{dcolumn}
\usepackage{bm}
\usepackage[utf8]{inputenc}
\usepackage{float}
\usepackage{amssymb}
\usepackage[T1]{fontenc}
\usepackage{color,soul}
\usepackage{mathtools}
\usepackage{epsfig}
\usepackage{color}
\usepackage{url}
\usepackage{siunitx}

\usepackage{mathptmx}

\begin{document}

\preprint{AIP/123-QED}

\title[Multi-bit MRAM storage cells utilizing serially connected perpendicular magnetic tunnel junctions]{Multi-bit MRAM storage cells utilizing serially connected perpendicular magnetic tunnel junctions}

\author{Piotr Rzeszut}
\email{piotrva@agh.edu.pl}
\author{Witold Skowroński}
\author{Sławomir Zi\ifmmode \mbox{\k{e}}\else \k{e}\fi{}tek}
\affiliation{AGH University of Science and Technology, Department of Electronics,\\ Al. Mickiewicza 30, 30-059 Krak\'{o}w, Poland}%
\author{Jerzy Wrona}
\affiliation{Singulus technologies, Kahl am Main, 63796, Germany}%
\author{Tomasz Stobiecki}
\affiliation{AGH University of Science and Technology, Department of Electronics,\\ Al. Mickiewicza 30, 30-059 Krak\'{o}w, Poland}%
\affiliation{AGH University of Science and Technology, Faculty of Physics and Applied Computer Science,\\Al. Mickiewicza 30, 30-059 Krak\'{o}w, Poland}%

\date{\today}

\begin{abstract}
Serial connection of multiple memory cells using perpendicular magnetic tunnel junctions (pMTJ) is proposed as a way to increase magnetic random access memory (MRAM) storage density. Multi-bit storage element is designed using pMTJs fabricated on a single wafer stack, with a serial connections realized using top-to-bottom vias. Tunneling magnetoreistance effect above \SI{130}{\percent}, current induced magnetization switching in zero external magnetic field and stability diagram analysis of single, two-bit and three-bit cells are presented together with thermal stability. The proposed design is easy to manufacture and can lead to increase capacity of future MRAM devices.
\end{abstract}

\maketitle

\section{\label{sec:Introduction}Introduction}
Spin transfer torque magnetoresistive random access memories (STT-MRAM) have numerous advantages over existing storage technologies, including theoretically unlimited endurance, high read and write speeds and ionizing-cosmic-radiation resistance\cite{kent2015new, dieny2010spin}. However, state-of-the-art memories have limited capacity due to the fact, that current density needed to switch a cell (typically made of a single magnetic tunnel junction) requires relatively large transistors\cite{kawahara20072mb}. Such an obstacle can be overcome using the architecture that incorporates multibit cell driven by a single transistor.

To date, very few practical implementations of multi-bit MRAM cells have been presented\cite{jeong1999three, raymenants2018chain}. This is mainly due to the fact, that efforts were made to produce a single storage element capable of being stable in more than two states, or to produce multiple storage elements on the top of each other\cite{ju2006multibit, ishigaki2010multi, lequeux2016magnetic}. Both of these approaches are very challenging for the process of manufacture.

In this work an alternative approach is proposed – perpendicular magnetic tunnel junctions (pMTJs)\cite{ikeda2010perpendicular} are connected electrically in series and a multi-state behaviour is observed, that leads to a multi-bit storage capability. Theoretical explanation as well as experimental results (including working three-bit cell) are presented. In addition, such an approach can be implemented to design and fabricate an artificial synapse for neuromorphic computing scheme\cite{zhang2016all, torrejon2017neuromorphic, lequeux2016magnetic, sung2018perspective, sulymenko2018ultra, fukami2018perspective}.

\section{\label{sec:Principles}Principles of operation}
\begin{figure}[h!]
\begin{center}
	\includegraphics[width=\columnwidth]{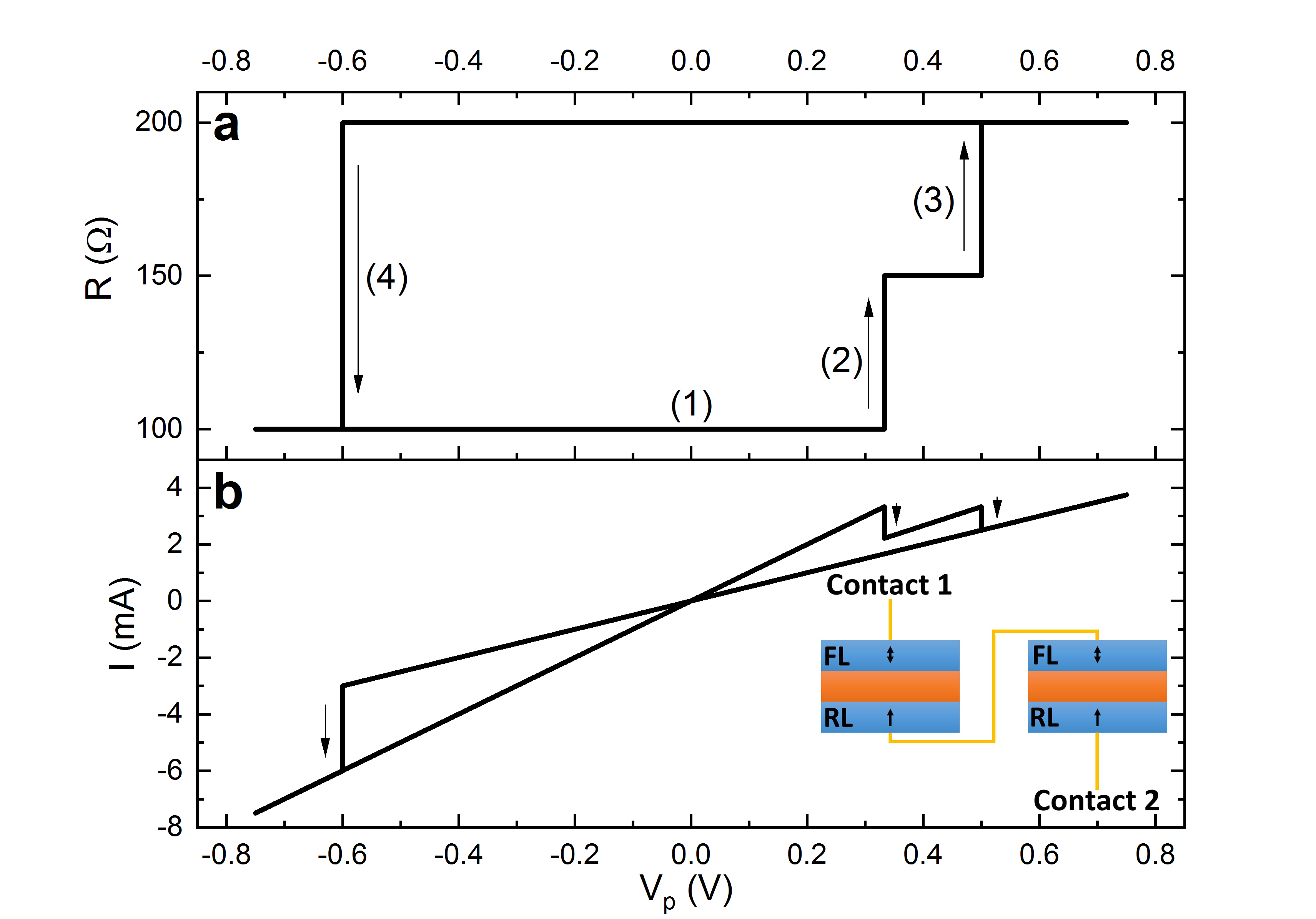}
	\caption{Theoretically predicted a) resistance and b) current versus voltage applied to a storage cell consisting of two serially connected pMTJs. Inset: Schematic of serial connection of two pMTJs.}
\label{fig:theory_prediction}
	\end{center}
\end{figure}

The discussed pMTJs consist of a top free layer (FL), MgO tunnel barrier and a bottom reference layer (RL), which is magnetically pinned to the synthetic ferromagnet (SyF) \cite{worledge2011spin}. In the proposed serial connection of pMTJs in a storage cell, the top contact of the first element is connected to the bottom contact of the next element (head-to-tail), as shown in the inset of Fig. \ref{fig:theory_prediction}. This results in the charge current flowing through all the cells involved in the same direction. Connections can be made using metallization and vias or any other suitable technique. 

The behaviour of the presented arrangement of storage elements can be predicted by analysing characteristics of two pMTJs connected (Fig. \ref{fig:theory_prediction}). If both elements are in the parallel (P) state, the lowest resistance is observed (1). When a positive voltage is applied (which corresponds to the current flow that favours anti-parallel (AP) state), the current increases, until it reaches critical value, which results in the current induced magnetization switching (CIMS) (2). As one of the elements switches to the AP state, with constant voltage applied, the current decreases. This prevents the remaining element of the cell from switching, as the current drops below the critical value. By further increasing the voltage, the critical current is reached again, and the second pMTJ switches to the AP state (3).

By reversing the current polarization, the switching to the P state is achieved. In this case, as soon as the critical current is reached, one of the elements switches to the P state (4). With constant voltage applied, the current rises above the critical value, causing the other element to switch to P state. 

The above mechanism works also for more than two elements, and similar reasoning can be carried out. For the serial pMTJs connection utilizing the presented mechanism, $N+1$ stable resistance states would be observed for $N$ elements connected, resulting in storage ability of $log_2(N+1)$ bits. This is because there is no possibility to individually determine states of all incorporated storage elements, as ideally they are characterized by the same resistance - only the number of elements in P and AP state may be determined, based on the two-point resistance measurement.

\begin{figure}[h!]
\begin{center}
	\includegraphics[width=\columnwidth]{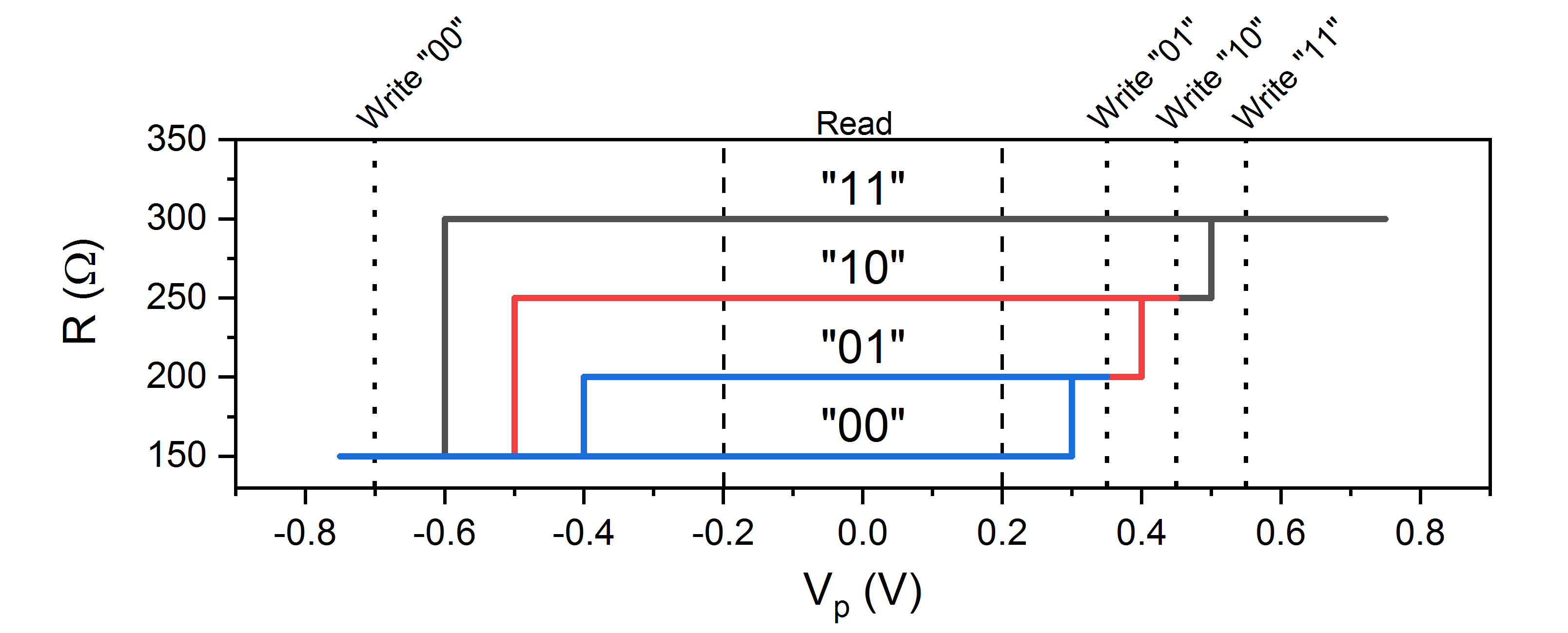}
	\caption{Theoretically predicted resistance versus voltage applied to the proposed two-bit cell. Possible mapping between resistance and binary value as well as proposed voltages to write and read the cell are presented on the plot. Different colours represent the behaviour of the cell after different writing voltages application.}
\label{fig:theory_prediction3}
	\end{center}
\end{figure}

The storage cell capable of storing two bits of data would, therefore, consist of three serially connected storage elements. The predicted resistance vs. voltage characteristics of such a storage cell are presented in Fig. \ref{fig:theory_prediction3}. Voltages for writing different states, as well as reading the cell can be defined based on the characteristics obtained for a single pMTJ. Note, that in the proposed cell configuration, writing smaller bit value (smaller resistance) than the existing state requires clearing the state to "00" (the lowest resistance) and writing a new value.

A simillar solution was suggested by Raymenants et al. in Ref. \onlinecite{raymenants2018chain}; however, with a different arrangement of subsequent elements, which are connected in opposite directions (head-to-head and tail-to-tail). Such a multi-level cell, though, needs application of variable external magnetic field to have an ability to be written with any desired state, what is not a case for our design, where read-write process is much simpler. On the other hand, our solution has by design a limited number of stable states to $N+1$.

Similar mechanism was also suggested by Zhang et. al.\cite{zhang2016all}, with elements fabricated on top of each other; however, due to experimental difficulties only two working elements connected in series were fabricated.

In our work these drawbacks of other designs are eliminated, and the presented arrangement is ready to be manufactured using unmodified fabrication process.

\section{\label{sec:Experimental}Experiment}
\begin{figure}[h!]
\begin{center}
	\includegraphics[width=\columnwidth]{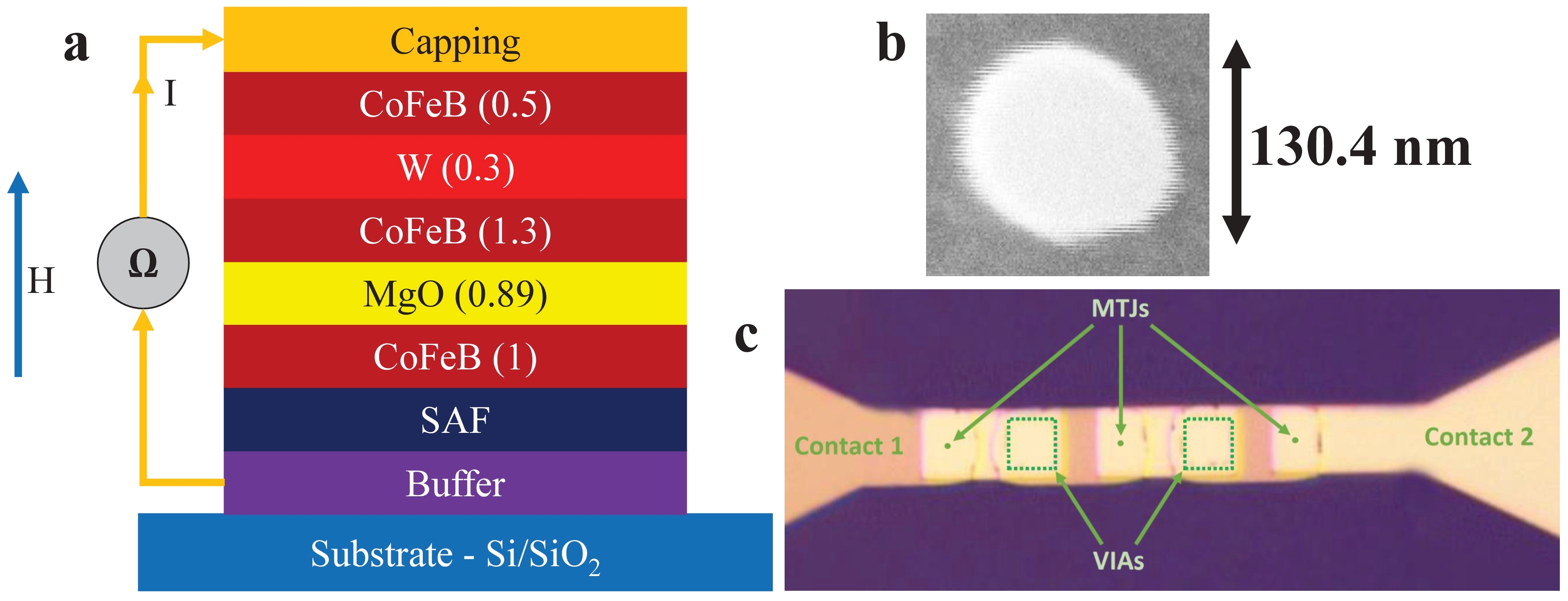}
	\caption{a) pMTJ layer structure, $SAF$ denotes synthetic antiferromagnet, $H$ - external magnetic field direction, $I$ - current direction, $\Omega$ - sourcemeter or cell driving circuit. b) Scanning electron microscope image of a single pMTJ. c) Micrograph of three serially connected pMTJs.}
\label{fig:element_drawing}
	\end{center}
\end{figure}

Multilayer of the following structure: buffer / Co(0.5)-Pt(0.2) based SyF / W(0.25) / CoFeB(1) / MgO(0.89) / CoFeB(1.3) / W(0.3) / CoFeB(0.5) / MgO(0.75) / capping layers (thickness in \SI{}{\nano\meter}) patterned into pillars of around \SI{130}{\nano\meter} diameter were used as pMTJ basic cell. The details of the deposition and fabrication processes are presented in Refs. \onlinecite{sato2012cofeb, skowronski2017understanding, skowronski2018influence}. Elements were equipped \SI[product-units = power]{100 x 100}{\micro\meter} Al(20)/Au(30) contact pads that enable both individual pMTJ characterization as well as measurement of the elements connected in series forming a multi-bit cell. The schematics of the multilayer stack, fabricated pMTJ pillar and a micrograph of a two-bit (three pMTJs in series) cell are presented in Fig. \ref{fig:element_drawing}. In order to determine an ability of a single element to act as a memory device, two types of characterization were performed: a stability diagram\cite{skowronski2017understanding} and thermal stability\cite{sato2012cofeb} measurements. 

The stability diagram was determined as follows: pMTJ resistance ($R$) versus voltage pulse amplitude ($V_p$) measurements were repeated with different external magnetic field ($H$) applied. Voltage pulse length was set to \SI{10}{\milli\second}. Each point on the stability diagram corresponds to the transition from the P to AP state or AP to P state, depending on the initial magnetization configuration. The thermal stability was determined from the $R$ vs. $H$ measurement repeated around hundred times with a fixed sweep rate of around \SI[per-mode=symbol]{80}{\ampere\per\metre\per\second}.

\section{\label{sec:Results}Results and discussion}
\subsection{\label{sec:Single}Single pMTJ characterization}

\begin{figure}[h!]
\begin{center}
	\includegraphics[width=\columnwidth]{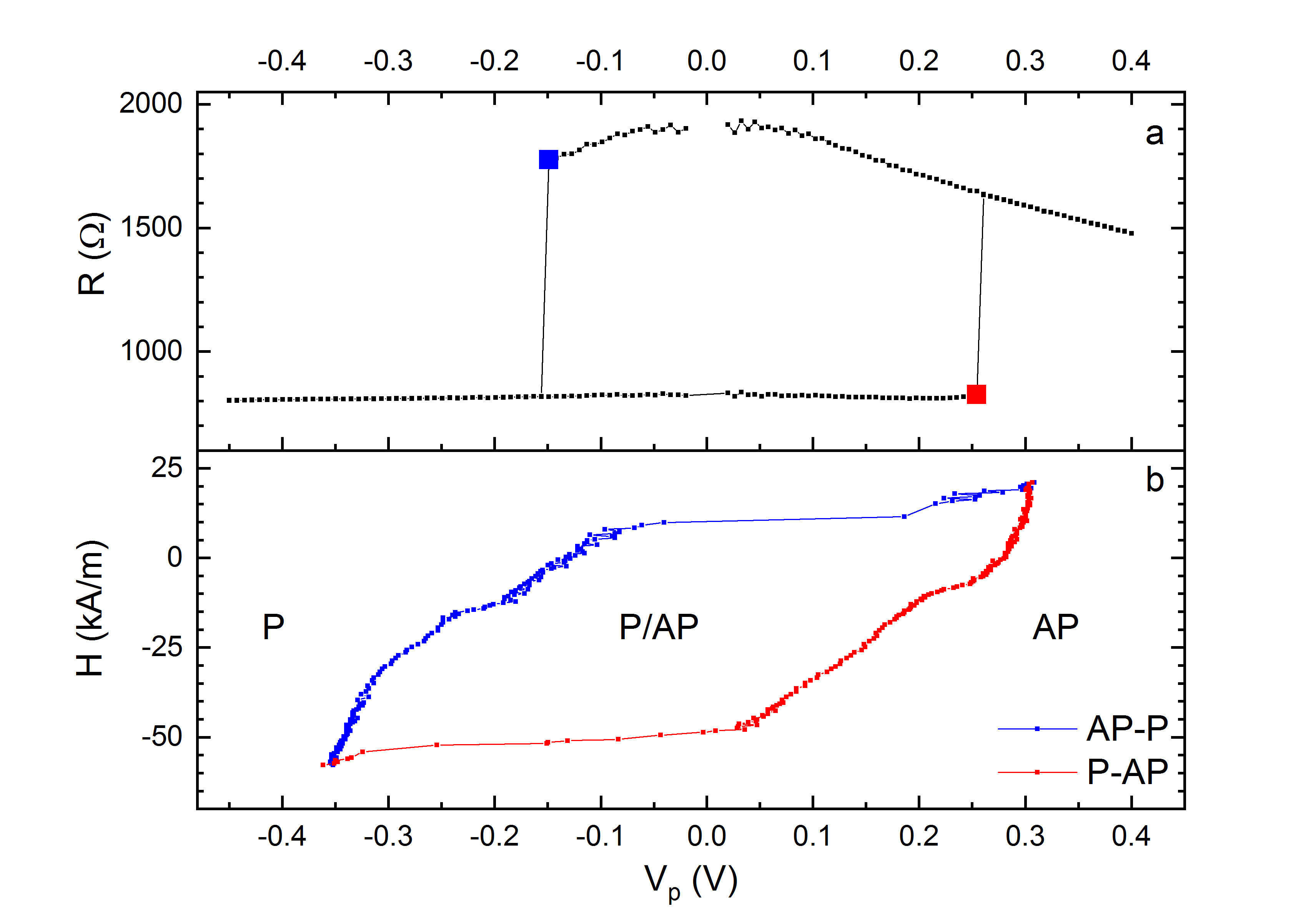}
	\caption{a) Representative $R$-$V$ loop of a single pMTJ, with switching voltages from P to AP (red) and from AP to P (blue) marked using big squares, measured without external magnetic field. b) A stability diagram with marked regions where P, AP or both of the states are stable.}
\label{fig:eye_diagram}
	\end{center}
\end{figure}

An example of the $R(V_p)$ loop and the stability diagram are presented in Fig. \ref{fig:eye_diagram}. The TMR ratio of \SI{135}{\percent} and resistance area (RA) product of \SI{21.6}{\ohm\micro\metre\squared} were measured. These values, however, are influenced by series resistance of vias and contacts, which could not be eliminated due to two-wire measurement; in fact, the TMR ratio of the element is higher\cite{skowronski2018influence}. In the absence of an external magnetic field, the P to AP transition occurs for the voltage of around \SI{0.25}{\volt} (corresponding to the critical current density of \SI[per-mode=symbol]{2.00}{\mega\ampere\per\centi\metre\squared}, whereas, the AP to P switching is measured for \SI[parse-numbers = false, number-math-rm = \ensuremath, per-mode=symbol]{V_p = -0.15}{\volt} (corresponding to \SI[parse-numbers = false, number-math-rm = \ensuremath, per-mode=symbol]{J_{crit} = -0.54}{\mega\ampere\per\centi\metre\squared}). Multiple $R(H)$ measurements allowed to obtain the switching propability versus H using the analysis described in Ref. \onlinecite{sato2012cofeb} with the following equation

\begin{equation} \label{eq:TMRsato}
	P(\tau) = 1 - exp\left [ -\frac{\tau}{\tau_0}exp\left \{ -\Delta\left ( 1-\frac{|H-H_s|}{H_k^{eff}} \right ) \right \} \right ]
\end{equation}

In Eq. \ref{eq:TMRsato} $\tau$ denotes the magnetic field step duration (in this study \SI[parse-numbers = false, number-math-rm = \ensuremath, per-mode=symbol]{\tau = 1}{\second}), $\tau_0$ the inverse of the attempt frequency (in this work assumed to be \SI{1}{\nano\second}), $\Delta$ the thermal stability, $H_s$ shift field and $H_k^\mathrm{eff}$ denotes the effective magnetic anisotropy field.

\begin{figure}[h!]
\begin{center}
	\includegraphics[width=\columnwidth]{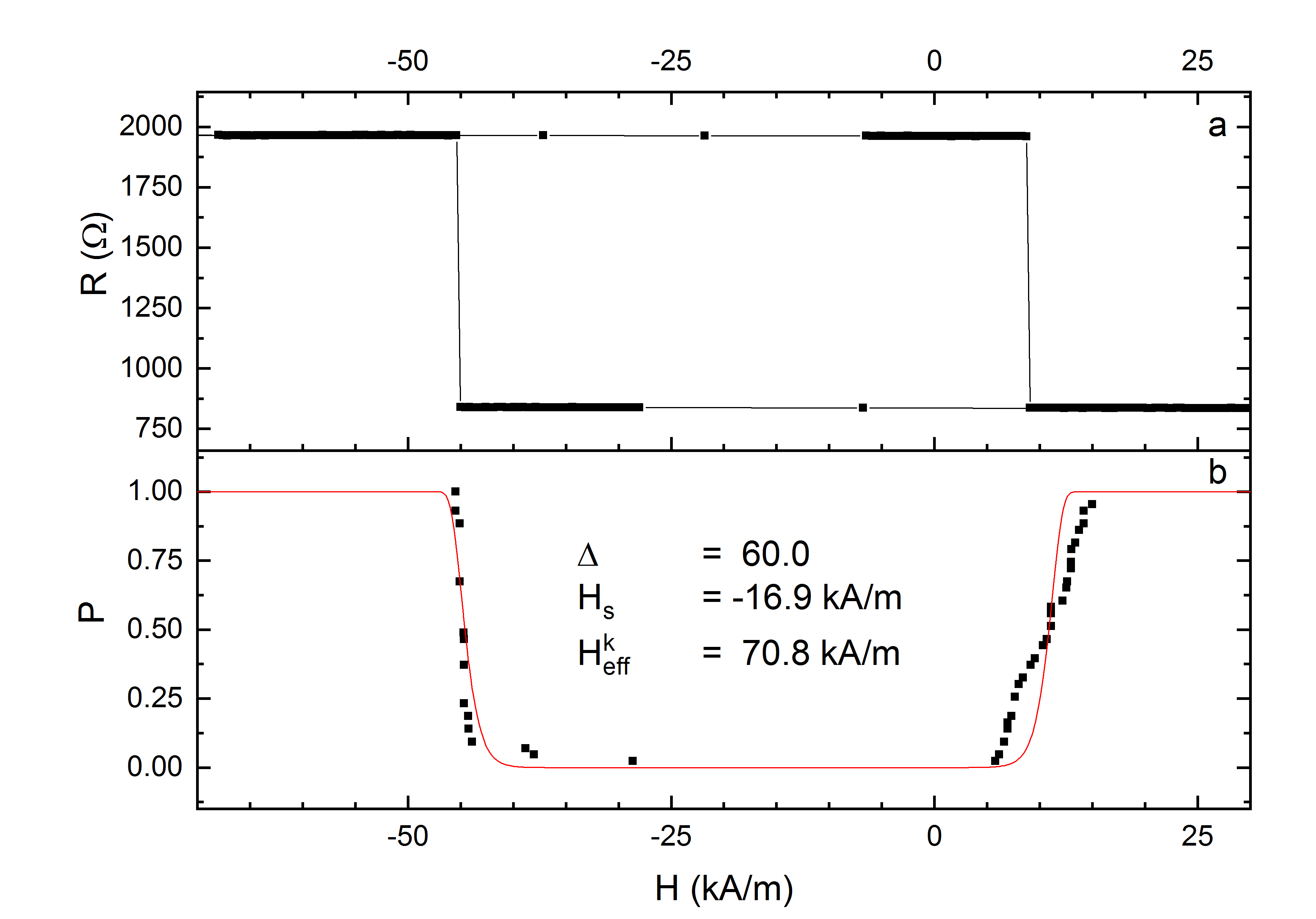}
	\caption{a) Representative $R$-$H$ loop. b) Calculated switching probability (black points) and theoretical fit based on Eq. \ref{eq:TMRsato}.}
\label{fig:sato}
	\end{center}
\end{figure} 

\begin{figure}[h!]
\begin{center}
	\includegraphics[width=\columnwidth]{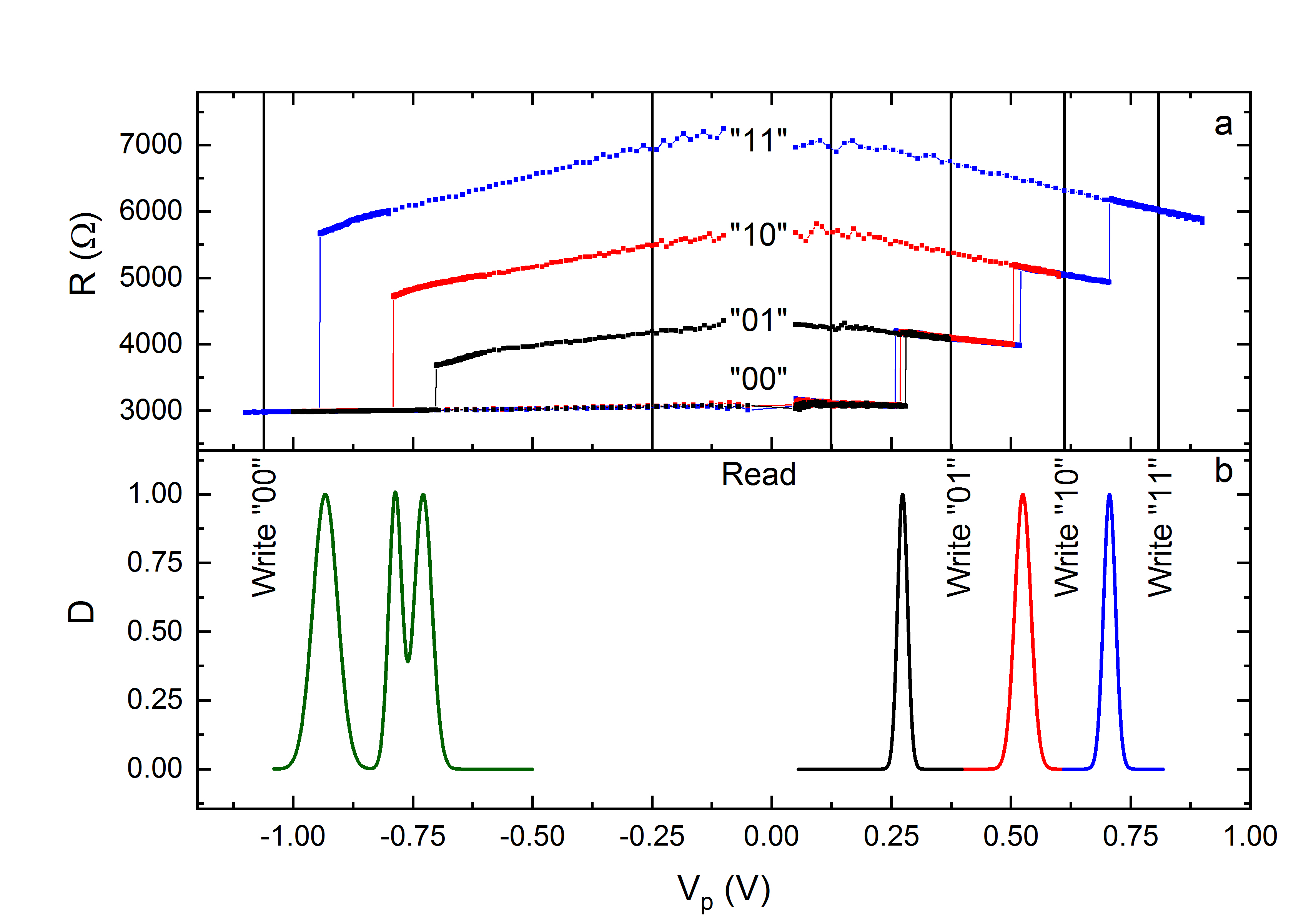}
	\caption{a) CIMS measurement of two-bit memory cell. The proposed binary coding is also presented. b) Switching voltages distribution, with regions for writing, and safe readout presented.}
\label{fig:two_bit_cims}
	\end{center}
\end{figure} 

The best fit of Eq. \ref{eq:TMRsato} to the experimental switching probability resulted in \SI[parse-numbers = false, number-math-rm = \ensuremath, per-mode=symbol]{\Delta = 60}{}, which, together with a capability of the pMTJ of being stable in both P and AP states in the absence of an external magnetic field proves that the cell is suitable to be used as a memory device.

\subsection{\label{sec:Experimental_twobit}Two-bit storage cell}
Next, we move on to the two-bit cell consisting of three pMTJs connected in series.
$R(V_p)$ measurement of such a system is presented in Fig. \ref{fig:two_bit_cims}a - initially two-bit cell is in the low resistance state. Application of the positive voltage of around \SI{0.37}{\volt} (corresponding to a single pMTJ switching from P to AP state) results in the transition to higher resistance state, which is denoted as "01". Further increase of voltage to around \SI{0.67}{\volt} causes a second pMTJ transition to a higher resistance state - thus "10" state is written. Finally, after application of \SI{0.80}{\volt}, all three pMTJs are in the AP state, which is denoted as "11" state. Negative voltage of \SI{-1.15}{\volt} switches all pMTJs back to P state.
The behaviour described in Sec. \ref{sec:Principles} was confirmed – four stable states can be defined and binary numbers can be assigned to them:
\begin{itemize}
\item all elements in AP state - 11
\item one element in P state and two in AP state - 10
\item two elements in AP state and one in P state - 01
\item all elements in P state - 00
\end{itemize}

By repeating $R$-$V$ measurement of two-bit cell around hundred times and calculating switching voltages distributions, writing voltages of particular states, as well as a region safe for reading the storage cell can be defined (Fig. \ref{fig:two_bit_cims}b).

\begin{figure}[h!]
\begin{center}
	\includegraphics[width=\columnwidth]{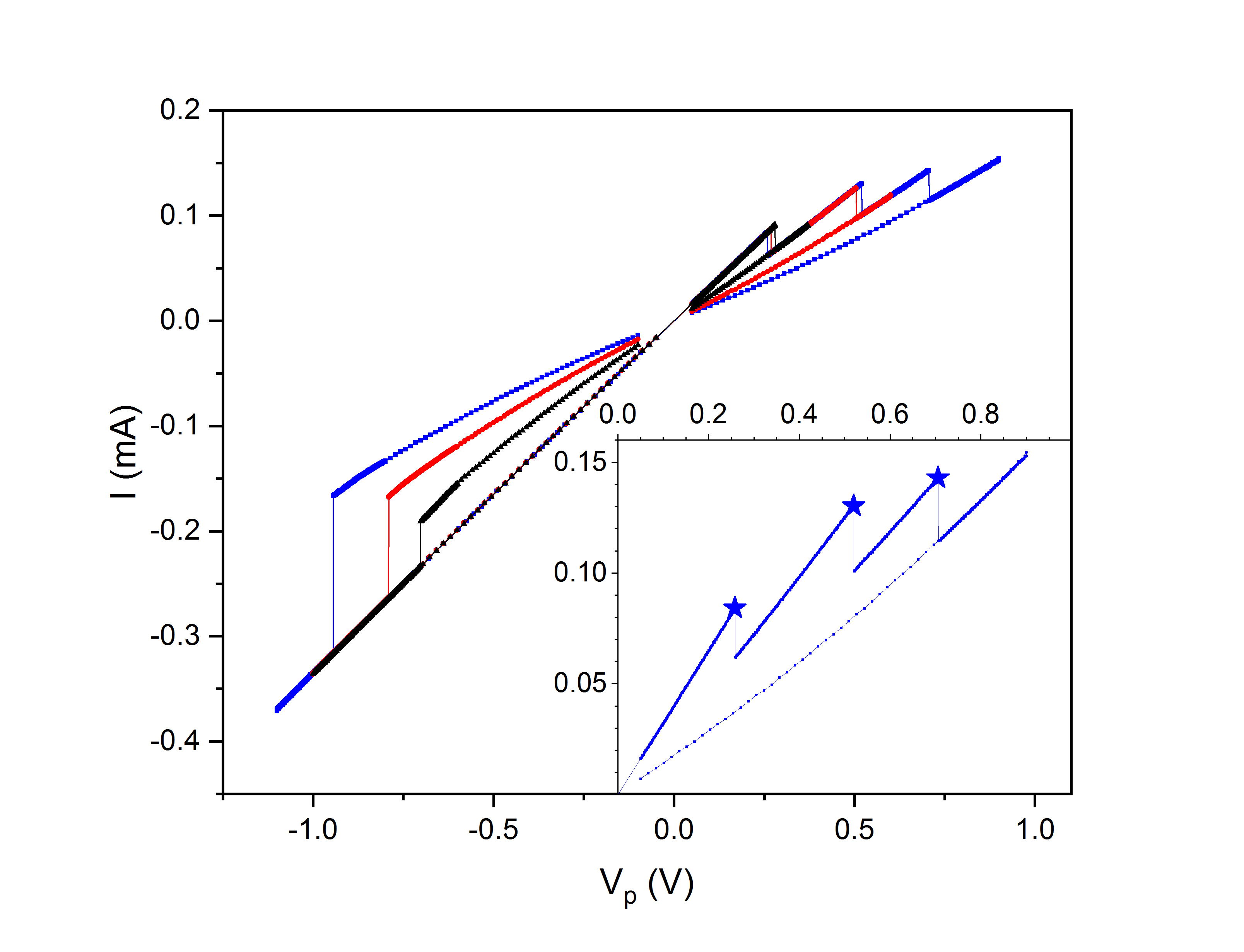}
	\caption{Current changes during $R$-$V$ measurement for storage cell constructed of three storage elements. Inset: a close up of current in the CIMS measurement for writing "11" cycle. Critical currents, causing subsequent elements to switch to the AP state are marked with stars.}
\label{fig:two_bit_current}
	\end{center}
\end{figure} 

The principle of operation, involving the current decreasing below the critical current after one element switching into AP state, was confirmed (Fig. \ref{fig:two_bit_current}). Due to non-ideal manufacturing process, critical currents of all incorporated elements are non-equal, but this has no adverse effect on the process (Fig. \ref{fig:two_bit_current} inset).

\subsection{\label{sec:Experimental_threebit}Three-bit storage cell}
\begin{figure}[h!]
\begin{center}
	\includegraphics[width=\columnwidth]{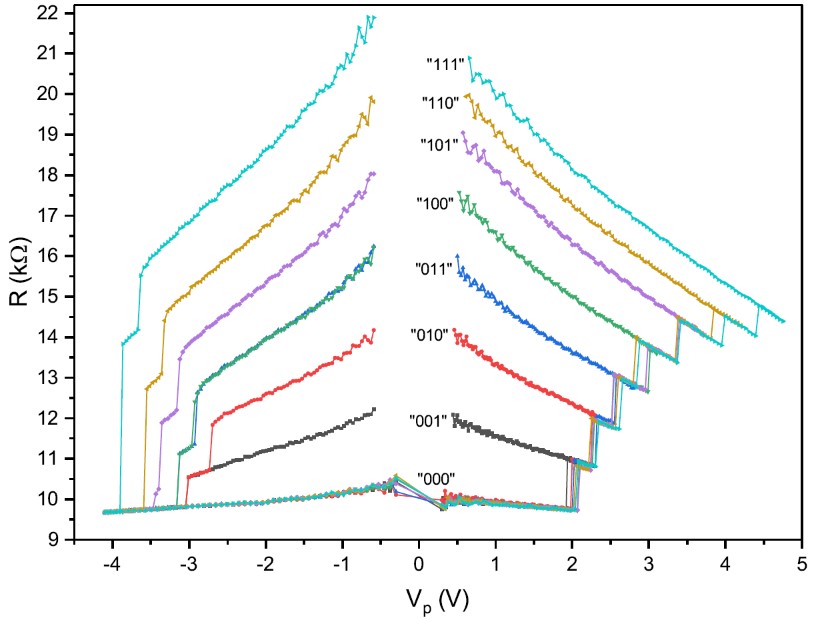}
	\caption{$R$-$V$ measurement of three-bit cell consisting of seven pMTJs connected in series. The proposed binary coding is presented.}
\label{fig:experiment_3bit}
	\end{center}
\end{figure}

Finally, the proof-of-concept of three-bit cell consisting of seven pMTJs connected in series is presented. As predicted, the cell exhibited eight stable states (Fig. \ref{fig:experiment_3bit}). Due to non-ideal fabrication process it was noted, that regions for writing voltages of some of the states are very narrow because of variation of the switching voltage (related with the switching current distribution). It is believed, that the behaviour is due to very similar critical currents of all incorporated elements. Also switching to "000" state was not ideal in the case, and needs further investigation.
Nonetheless, the proposed architecture is valid for a multiple pMTJ that form multi-bit memory cell. 

\section{\label{sec:summary_final}Summary}
In summary, we showed that multi-bit memory cell can be successfully implemented using serially connected pMTJs. State-of-the-art multilayer structure characterized by TMR of \SI{135}{\percent} and RA of \SI{21.6}{\ohm\micro\metre\squared} was used to design two- and three-bit MRAM cells. 
The developed method of fabrication and driving multi-bit non-volatile storage elements is a significant improvement in MRAM technology, as it allows to store more data using the same area of the memory. This may be achieved by driving a multi-bit storage cell using a single transistor rated for the same current, as a single storage element (the critical current remains the same for any number of serially connected elements). Also, the fabrication process does not require significant changes compared to single storage element fabrication.
In addition, the proposed solution maybe utilized in neuromorphic computing scheme as a multi-state non-volatile memory block.

\section*{Acknowledgement}
This work is supported by the Polish Ministry of Science and Higher Education Diamond Grant No. 0048/DIA/2017/46 and the Polish National Centre for Research and Development Grant No. LIDER/467/L-6/14/NCBR/2015.

TS acknowledges the  SPINORBITRONICS project through the National Science
Centre Poland under grant No. 216/23/B/ST3/01430.

The nano-fabrication process was performed at Academic Centre for Materials and Nanotechnology (ACMiN) of AGH University of Science and Technology.

\bibliography{bibliography}

\end{document}